# Accelerating Amorphous Alloy Discovery: Data-Driven Property Prediction via General-Purpose Machine Learning Interatomic Potential

Xuhe Gong [a,b], Hengbo Zhao [a], Xiao Fu [b], Jingchen Lian [b], Qifan Yang [b],

Ran Li [a,*], Ruijuan Xiao [b,*], Tao Zhang [a], Hong Li [b]

[a] *School of Materials Science and Engineering, Key Laboratory of Aerospace Materials and Performance (Ministry of Education), Beihang University, Beijing 100191, China;*

[b] *Institute of Physics, Chinese Academy of Sciences, Beijing 100190, China*

*E-mail: liran@buaa.edu.cn (R. Li), rjxiao@iphy.ac.cn (R. Xiao).

## Abstract

While traditional trial-and-error methods for designing amorphous alloys are costly and inefficient, machine learning approaches based solely on composition lack critical atomic structural information. Machine learning interatomic potentials, trained on data from first-principles calculations, offer a powerful alternative by efficiently approximating the complex three-dimensional potential energy surface with near-DFT accuracy. In this work, we develop a general-purpose machine learning interatomic potential for amorphous alloys by using a dataset comprising 20400 configurations across representative binary and ternary amorphous alloys systems. The model demonstrates excellent predictive performance on an independent test set, with a mean absolute error of 5.06 meV/atom for energy and 128.51 meV/Å for forces. Through extensive validation, the model is shown to reliably capture the trends in macroscopic property variations such as density, Young's modulus and glass transition temperature across both the original training systems and the compositionally modified systems derived from them. It can be directly applied to composition-property mapping of amorphous alloys. Furthermore, the developed interatomic potential enables access to the atomic structures of amorphous alloys, allowing for microscopic analysis and interpretation of experimental results, particularly those deviating from empirical trends.

This work breaks the long-standing computational bottleneck in amorphous alloys research by developing the first general-purpose machine learning interatomic potential for amorphous alloy systems. The resulting framework provides a robust foundation for data-driven design and high-throughput composition screening in a field previously constrained by traditional simulation limitations.

## 1 Introduction

The conventional compositional design of amorphous alloys has traditionally been guided by established empirical criteria, such as eutectic point criterion[1], the confusion principle[2] and Inoue's three empirical rules[3], ultimately relying on experimental "trial-and-error" methods. While these methods have yielded significant research progress over the years[4–6], a more efficient and cost-effective approach is still critically needed, especially when dealing with the vast candidate space of multi-component systems. With the accumulation of raw experimental data and the development of computer technology, establishing machine learning models based on experimental dataset has become a new method for material development and mechanism exploration[7–9]. However, in the field of amorphous alloys, model performance is hindered by the imbalanced distribution of experimental data[10], and furthermore, descriptors solely derived from elemental compositions lack atomic structural information, limiting interpretability at the microscopic level.

To achieve high-precision simulations that provide both the prediction and mechanistic insights into the properties of amorphous alloys at the atomic scale, first-principles calculations based on density functional theory (DFT) are commonly used to validate experimental results and investigate their underlying mechanisms[11–13]. However, despite its high accuracy, DFT simulations are typically limited to systems of around $10^2$ atoms due to their high computational cost. The spatial and temporal scales accessible to DFT differ from experimental scales by several, or even over ten, orders of magnitude. It is insufficient to use periodic repeating units of around 1 nm to capture the isotropic and medium-range-order information of amorphous alloys. To perform large-scale molecular dynamics simulations of amorphous alloys, researchers commonly employ empirical potentials, such as the embedded atom method (EAM) potential, which is frequently applied to the CuZr-based system[14–16]. Nevertheless, these empirical potentials are often limited in transferability and fall short of the theoretical accuracy of first-principles calculations.

In recent years, machine learning interatomic potentials (MLIPs) have emerged as a powerful tool for materials design and mechanistic investigations[17–19], due to their exceptional accuracy and superior computational efficiency. MLIPs are developed by training models on datasets composed of atomic configurations, along with their associated energies and forces, allowing the models to approximate the three-dimensional potential energy surface. On the one hand, with the results of first principles calculation as the training set, fully trained MLIPs achieve accuracy comparable to that of first-principles methods. On the other hand, modeling atomic interactions using potential functions, rather than directly solving the Schrodinger equation, significantly reduces computational cost. Numerous pretrained MLIPs have been developed in recent years, including notable examples such as CHGNet[20], DPA-2[21], and NEP89[22]. These models have consistently demonstrated remarkable accuracy on test sets and computational efficiency in molecular dynamics simulations. However, most online first-principles databases are primarily focused on crystalline materials. Consequently, these pretrained models exhibit limited performance when applied to high-energy structures that are scarce in the datasets, such as transition states, non-equilibrium configurations, and disordered systems. For amorphous materials, some researchers have already developed their own datasets and corresponding potential functions. Representative examples include MLIPs for Li–Al–Cl–O[18], Zr–Cu–O[23], and Pd-Cu-Ni-P[24] amorphous systems, which have been used to study ion transport, mechanical performance, and thermal behavior at the atomic scale. However, these potential functions are principally developed for specific materials, targeting either amorphous lithium-containing compounds or specific compositions within amorphous alloys. While these models provide valuable insights in their respective domains, their applicability to a wider range of amorphous alloy systems remains limited. To date, a universally applicable interatomic potential for the diverse systems of amorphous alloys remains unavailable.

In this work, we propose a methodological framework for developing a general-purpose MLIP for amorphous alloys. Initially, we construct a DFT dataset comprising 20400 configurations and fine-tune the pretrained CHGNet to obtain a high-precision MLIP for amorphous alloys. The resulting MLIP demonstrates outstanding robustness in simulating 19 representative amorphous alloy systems involving 25 different chemical elements. Utilizing the developed potentials, we perform simulations by Large-scale Atomic/Molecular Massively Parallel Simulator[25] (LAMMPS) to evaluate macroscopic physical properties, including density, Young's modulus, and glass transition temperature across multiple amorphous alloy systems. Simulation results exhibit strong agreement with experimental data and successfully reproduce anomalies observed during our composition development in experiment that deviate from empirical trends, providing insights into their underlying microscopic mechanisms.

The MLIP developed in this work effectively captures the composition dependence of key physical properties in amorphous alloys. Traditionally, the intrinsically disordered structures of amorphous alloys, combined with the scarcity and uneven distribution of experimental data, have posed significant challenges to data-driven materials design. However, the framework established in this work, including a dataset encompassing multiple amorphous systems, a highly generalizable MLIP, and a simulation process for macroscopic property prediction, provides a solid foundation for constructing amorphous alloy databases and conducting high-throughput composition design. This approach provides new tools and perspectives for accelerating the development of amorphous materials.

## 2 Result

### 2.1 Dataset Construction

In this study, we select six binary amorphous alloy systems (Ca-Al, Pd-Si, Cu–Zr, Cu–Hf, Ni–Nb, and Ni–Ta) and thirteen ternary systems (Au–Cu–Si, Ag–Mg–Ca, Mg–Cu–Y, Ce–Al–Co, La–Al–Co, Pt–Ni–P, Pd–Ni–P, Cu–Zr–Al, Ti–Zr–Be, Fe–Nb–B, Co–Ta–B, Mo–Co–B, and Ir–Ni–Ta), comprising a total of 19 compositionally simple amorphous alloy systems as training targets for the model. These systems are strategically chosen for their foundational role in bulk metallic glass research and broad coverage of key physical properties observed across amorphous alloys. As illustrated in Figure 1(a), these systems incorporate elements commonly found in amorphous alloys, including typical compositionally simple bulk metallic glasses (BMGs) systems. These classical systems, celebrated for their foundational role in BMGs design, continue to enable high-performance variants[13–15] through elements substitutional strategy. As illustrated in Figure 1(b), these systems demonstrate significant variability in key properties, including glass-forming ability (GFA), glass transition temperature, Young's modulus, and mass density. To enhance the generalization capability within individual systems, compositions are selected at intervals of 20 atomic percent. As shown in Figure 1(c), four different compositions are used for each binary system and six for each ternary system. The structural volume is determined by equilibrating the external pressure during ab initio molecular dynamics (AIMD) simulations conducted at 300 K using Vienna Ab initio Simulation Package (VASP)[26]. Atomic configurations are extracted at regular intervals from ab initio molecular dynamics (AIMD) trajectories conducted under the isothermal–isochoric (NVT) ensemble at 2500 K for each composition. For each binary system, 800 configurations are sampled, while 1200 configurations are sampled for each ternary system, resulting in a DFT dataset consisting of 20400 configurations in total. Concurrently, an entirely independent test dataset comprising 5100 configurations is constructed using the same method. The detailed dataset construction process is outlined in Figure S1.

**2.2 Model Training**

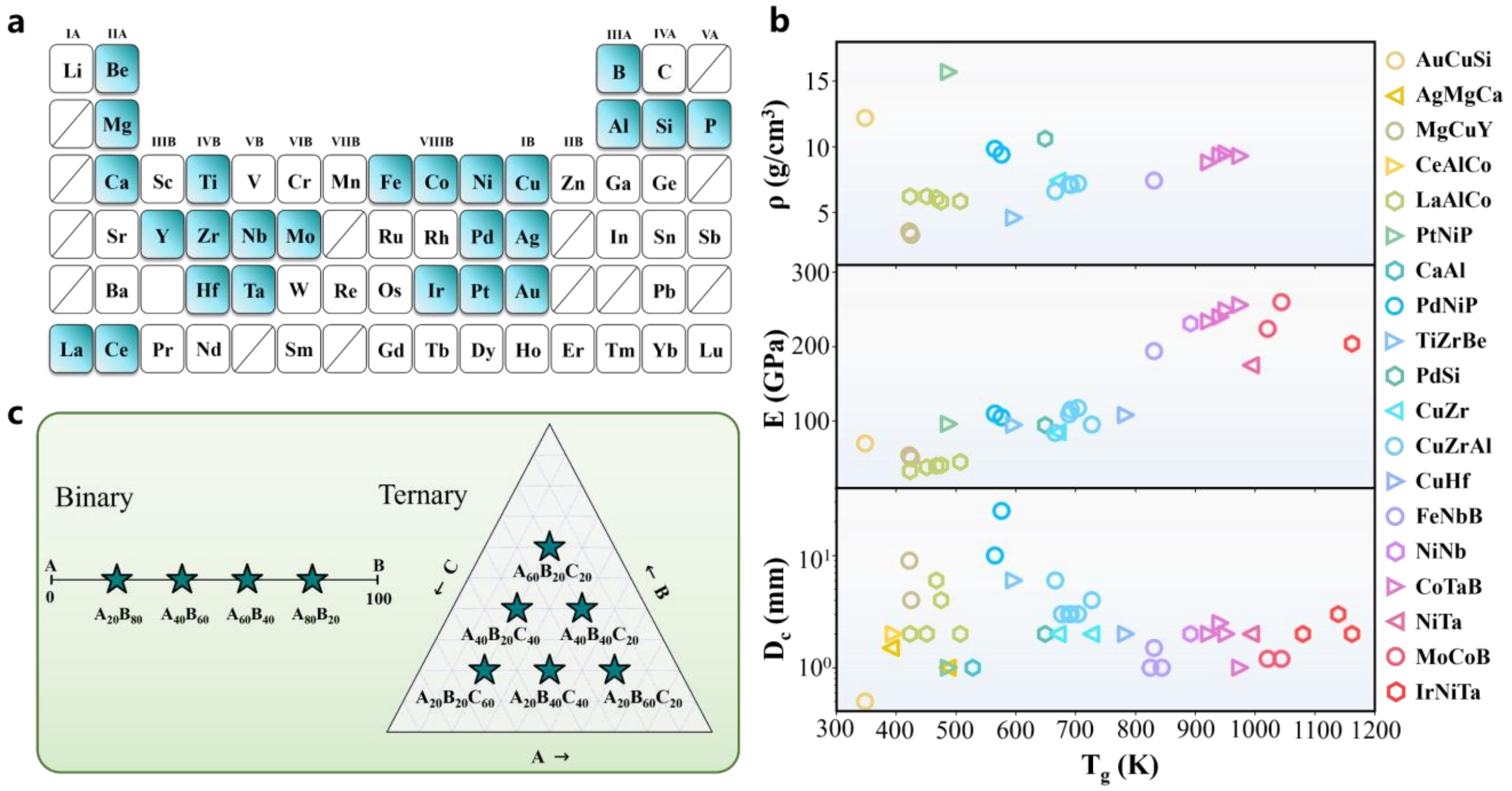

Figure 1 Dataset statistics and molecular dynamics calculation methods

(a) The distribution of the elements designed in the dataset in the periodic table, where the slash indicates that there are no reports of amorphous alloys containing this element for the time being

(b) The distribution of experimental data of amorphous alloy systems contained in the training set

(c) Schematic diagram of compositions selection for training sets of binary and ternary systems

Given the objective of developing a potential function applicable to multiple systems and a wide range of elements, we initiate model training based on the pretrained potential provided by CHGNet. This pretrained model is trained on 157955 configurations from the Materials Project (MP) database[27] and covers 89 elements from the periodic table. Since the database primarily consists of crystalline structures, the model exhibits considerable prediction errors when applied to the energies and forces of amorphous alloy configurations, as shown in Figure S2. Hence, additional training is needed to fine-tune the model specifically for accurate predictions of amorphous alloy structures.

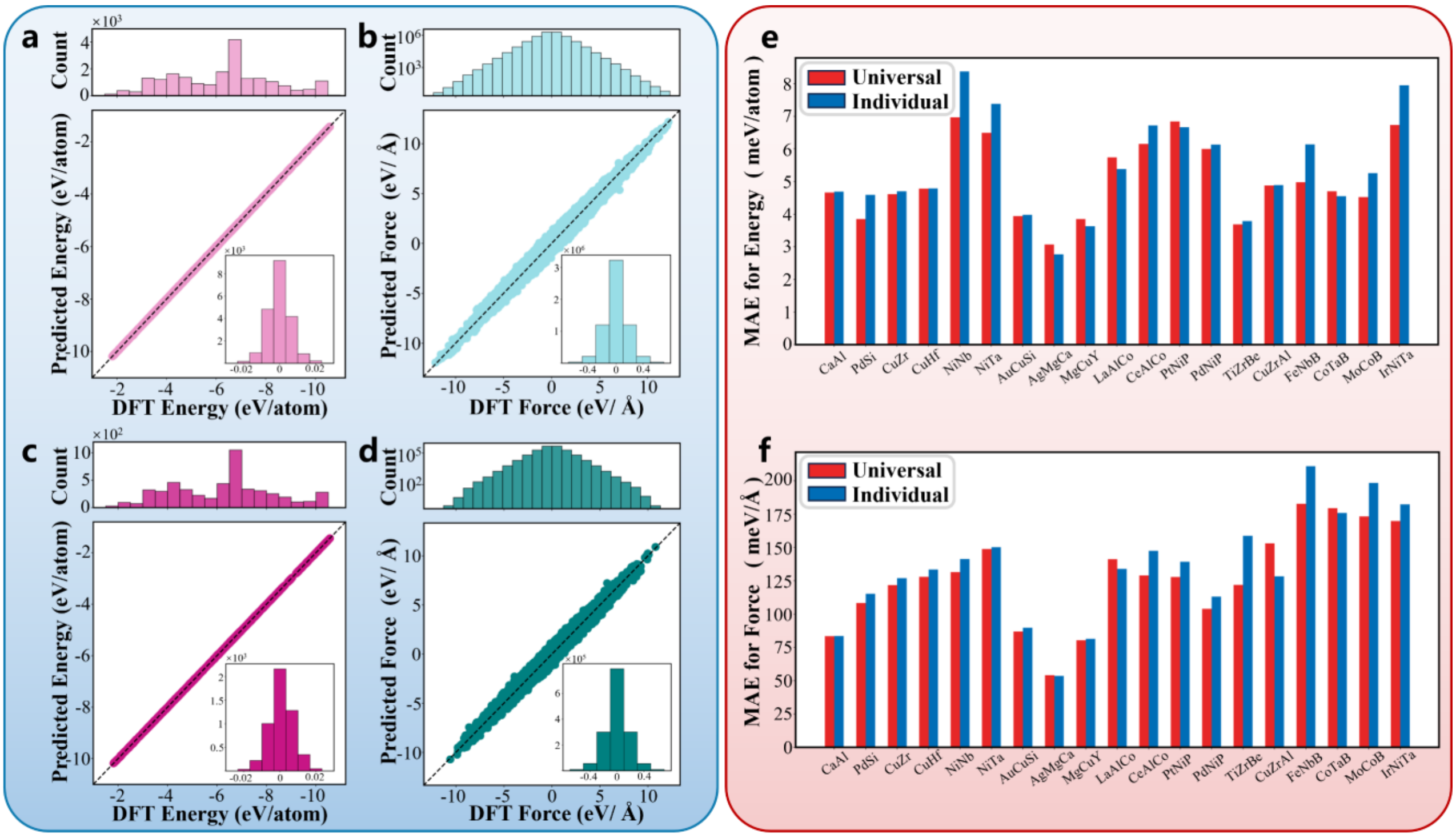

Figure 2 Model performance evaluation

Scatter plots in (a) and (b) show the comparison between DFT and predicted values for energies and forces in the training set, while (c) and (d) present the corresponding results for the test set. (e) and (f) compare the prediction performance across multiple systems under universal and individual training strategies for energies and forces, respectively.

To further evaluate the performance of the trained model, we analyze its predictions on the training set and the test set, as shown in Figures 2(a–d). Each figures compares the DFT-calculated values with model predictions, where the scatter points closely follow the black dashed line representing perfect agreement. The insets in the lower right of each subplot show the distribution of prediction errors, which exhibit typical Gaussian-like behavior centered around small absolute errors. The top subplots of figures display the distributions of energy and force values in the datasets, revealing that the training and test sets share statistically consistent distributions. This confirms the effectiveness of our test set sampling strategy and ensures its representativeness. Overall, the trained potential demonstrates strong fitting accuracy and generalization ability across the dataset, achieving a final energy prediction error of 5.06 meV/atom and a force error of 128.51 meV/Å.

As previously discussed, molecular dynamics simulations for amorphous alloys have predominantly focused on specific systems or compositions, and accordingly, the development of potential functions has typically been limited to individual alloys. In this study, the established potential function model demonstrates strong predictive performance across 19 distinct systems. We further compared different training strategies, as illustrated in Figure 2(e-f). The red bars represent the model trained on data from all alloy systems jointly, referred to as the universal model, while the blue bars correspond to models trained separately for each individual system, referred to as individual models. As shown in Figure 2(e), the energy prediction errors on the test set for each system are presented as bar charts. The universal model generally achieves lower prediction errors, with particularly significant improvements observed in systems such as Ni–Nb, Ni–Ta, Fe–Nb–B, Ti-Cu-Be, Mo–Co–B, and Ir–Ni–Ta. Similarly, Figure 2(f) reveals a clear advantage of the universal model over the individual models in predicting atomic forces across different systems. Notable reductions in force error are observed in systems including Fe–Nb–B, Mo–Co–B, and Ir–Ni–Ta.

These results demonstrate that simultaneously training the model on multiple systems does not compromise its performance, despite the increased complexity and size of the dataset. Previous studies often require the development of separate empirical or MLIPs for each individual system, which limits scalability and hinders cross-system applications. Conventional general-purpose potentials, such as the Lennard-Jones potential, exhibit inherent limitations in accuracy for modeling metallic glasses[28]. The potential developed in this work achieves near-DFT accuracy in predicting energies and forces across all systems included in the training set, which enables researchers to conduct cross-system molecular dynamics simulations using a unified potential. Furthermore, since CHGNet supports continued training on new datasets, the potential developed in this study can serve as a pretrained model for future simulations of other amorphous alloy research, reducing training costs and providing an efficient starting point for preliminary screening.

**2.3 Prediction of amorphous alloy macroscopic properties**

For amorphous alloys, density, Young's modulus, and glass transition temperature are essential properties for linking composition to performance and enabling materials screening. However, the intrinsic structural disorder and potential high-dimensional composition space of amorphous alloys continue to pose significant challenges for their systematic simulation. Experimental data, while valuable, are unevenly distributed across systems and compositions, making it difficult to construct a comprehensive database. Building on its accurate reproduction of atomic-scale energies and forces, the developed MLIP enables reliable prediction of key properties across diverse amorphous alloy systems, offering a valuable foundation for high-throughput calculations and composition design.

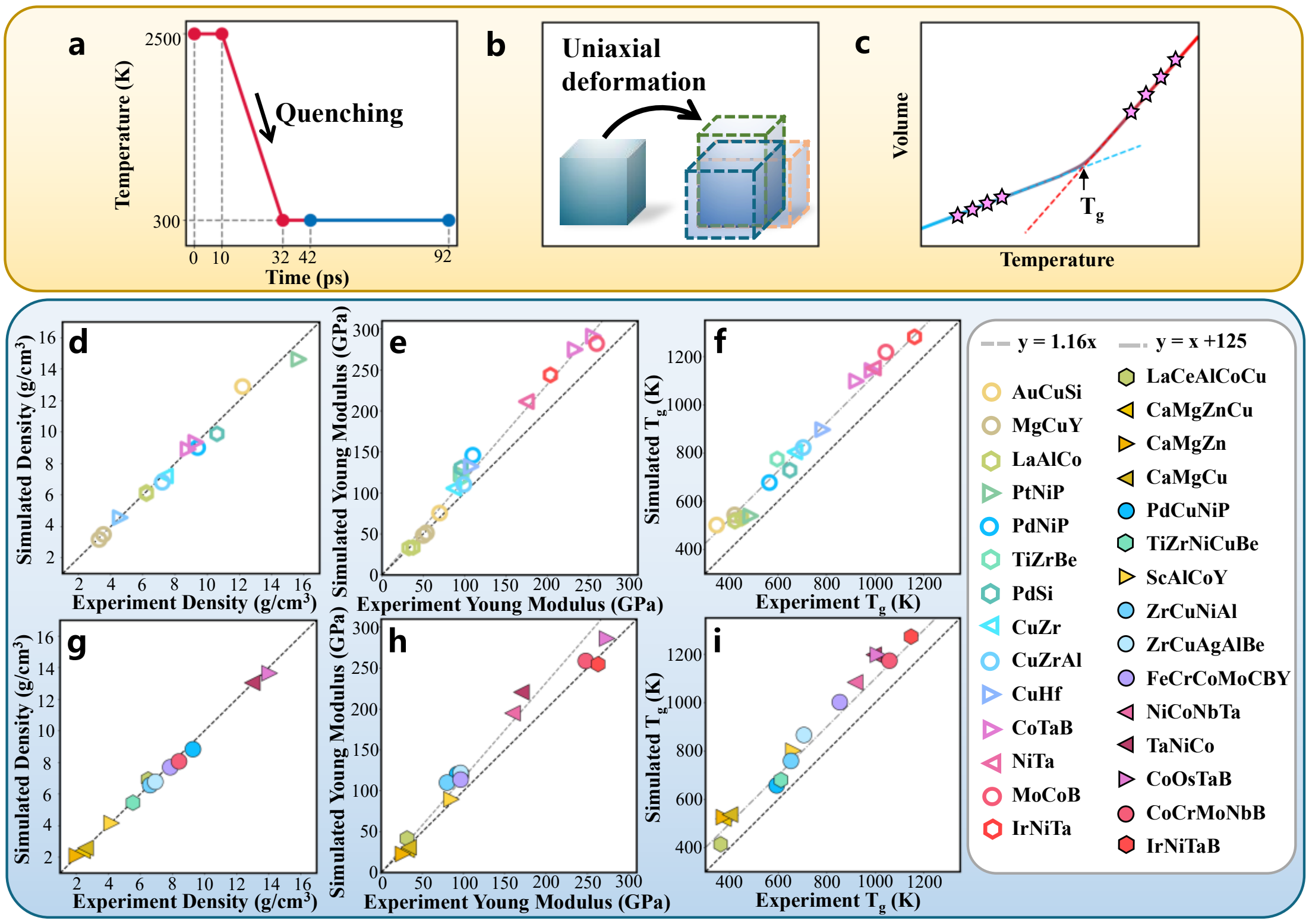

Figure 3 Simulation methods and results of macroscopic physical properties of amorphous alloys Schematic of the method for (a) amorphous structure generation and density analysis, (b) Young's modulus calculation, and (c) glass transition temperature determination. (d–f) Comparisons between simulated and reported values of (d) density, (e) Young's modulus, and (f) glass transition temperature for compositions included in the training set. (g–i) Comparisons between simulated and reported values of (g) density, (h) Young's modulus, and (i) glass transition temperature for extrapolated compositions.

We select several representative systems and compare the simulated values with corresponding experimental values for density, Young's modulus and glass transition temperature as shown in Figure 3. Figure 3(a-c) illustrates the methodology employed in these simulations, in which all systems are simulated using structures containing 1000 atoms. In Figure 3(a), the generation of amorphous structures is achieved by simulating the rapid quenching process used in experiments. Specifically, the structure is melted at 2500 K for 10 ps and then rapidly cooled to room temperature after 22ps, followed by a relaxation period of 10 ps to obtain the corresponding amorphous structure. The blue segment represents the process of isothermal annealing at 300 K for 50 ps, during which the radial distribution function (RDF) of the amorphous structure is analyzed. As is shown in the subfigure, a split second-neighbor peak confirms the formation of an amorphous structure. The average density over the final 10 ps of the isothermal annealing process is calculated and taken as the simulated value for comparison with experimental data. Figure 3(b) illustrates the procedure for calculating the Young's modulus. Uniaxial tensile deformations of 2% are individually applied along the X, Y, and Z directions of the amorphous structure in separate simulations. The Young's modulus is determined by calculating the slope of each stress–strain curve within the 0.5%–2% strain range, and then averaging the results across the three directions. Figure 3(c) illustrates the temperature-dependent volume change in amorphous materials, showing linear expansion in both the solid and liquid states, but with different slopes. The glass transition temperature is determined by the intersection of these two linear regions. Following this approach, we perform isothermal–isobaric ensemble (NPT) molecular dynamics simulations at four temperatures significantly below and four temperatures well above the glass transition temperature, respectively. As with the 300 K simulations used to obtain density, the equilibrium volume at each temperature is calculated by averaging the volume over the final 10 ps of a 50 ps simulation. Linear fits to the data from each temperature range are used to determine their intersection point, which defines the simulated glass transition temperature.

We compare the simulation results with 17 reported experimental data for selected systems from the training set, as shown in Figure 3(d-f), with detailed values listed in

Table S1. As shown in Figure 3(d), the simulated densities obtained using the MLIP closely match experimental values, with a consistent trend and an average relative error of ~3% across multiple systems. The comparison between simulated Young's modulus and experimental values is shown in Figure 3(e). We find that across several distinct amorphous alloy systems, the simulated Young's modulus is generally higher than the values reported in experiments. The average relative error is approximately 16%, as the scattered points cluster around the gray dashed line representing $y = 1.16x$. This deviation arises from multiple sources. On one hand, although MLIPs enable simulations at larger spatial scales than traditional first-principles methods, the simulated systems still differ from experimental systems by several orders of magnitude. On the other hand, reported experimental values of Young's modulus vary depending on the measurement method. Common methods for measuring Young's modulus in amorphous alloys include ultrasonic resonance, nanoindentation, and uniaxial deformation. Owing to disparities in measurement principles, the deformation behavior of a given composition may differ depending on the measurement technique employed.

Figure 3(f) presents a comparison of the simulated glass transition temperatures with reported experimental values. The simulated values are systematically higher than the experimental values, with an average overestimation of approximately 125 K, as the scattered points cluster around the gray dashed line representing $y = x + 125$. The primary source of this discrepancy lies in the excessively high heating rate employed in the simulation. In experiments, the glass transition temperature of amorphous alloys is typically measured using differential scanning calorimetry (DSC), where it is determined from the intersection of the extrapolated baseline and the tangent to the endothermic peak. The heating rate is a critical parameter that significantly influences the measured glass transition temperature, with researchers typically selecting a very slow heating rate, such as 0.33 K/s, to ensure uniformity and accuracy of the results. Although MLIPs extend the accessible timescale compared to AIMD, the upper timescale limit of simulations using MLIPs remains several orders of magnitude smaller than that of experiments. Previous studies have demonstrated that the measured glass transition temperature increases with higher heating rates[29]. Aside from the spatial scale limitations previously discussed, the primary source of error arises from the considerably higher heating rates used in simulations compared to those in experiments.

The above results demonstrate that the developed potential captures the correct trends in macroscopic property variations within the systems covered by the training set. However, in recent advances in amorphous alloy, multi-component alloy design has emerged as a key strategy for achieving superior performance. To assess the transferability and practical value of the potential, we simulate 15 reported amorphous compositions outside the training set, as shown in Figure 3(g-i). These additional compositions are generated by making minor elemental substitutions to systems in the training set, which is a common strategy in the development of new amorphous alloys. Although some elements are present only in the pretraining dataset, substitutions with similar elements do not significantly alter the atomic structures, which are precisely captured by our MLIP. The MLIP developed in this work demonstrates satisfactory generalization capability for structures not included in the training set, as shown in Figure S3. Consequently, simulations based on our MLIP remain capable of capturing the correct trends in macroscopic property variations. Detailed values are provided in Table S2. The density in Figure 3(g) also maintains good consistency with experimental values, with an average relative error of approximately 2%. Similarly, as shown in Figures 3(h) and (i), the simulated Young's modulus and glass transition temperatures follow the same trends as the experimental data, distributed along the lines $y = 1.16x$ and $y = x+125$, respectively. Given the strong similarity in trends between simulated and experimental values, the observed deviations of approximately 16% or 125 K can be considered systematic errors inherent to the simulation methodology. After applying this compensatory correction, the adjusted values show good agreement with experimental results, as illustrated in Figure S4.

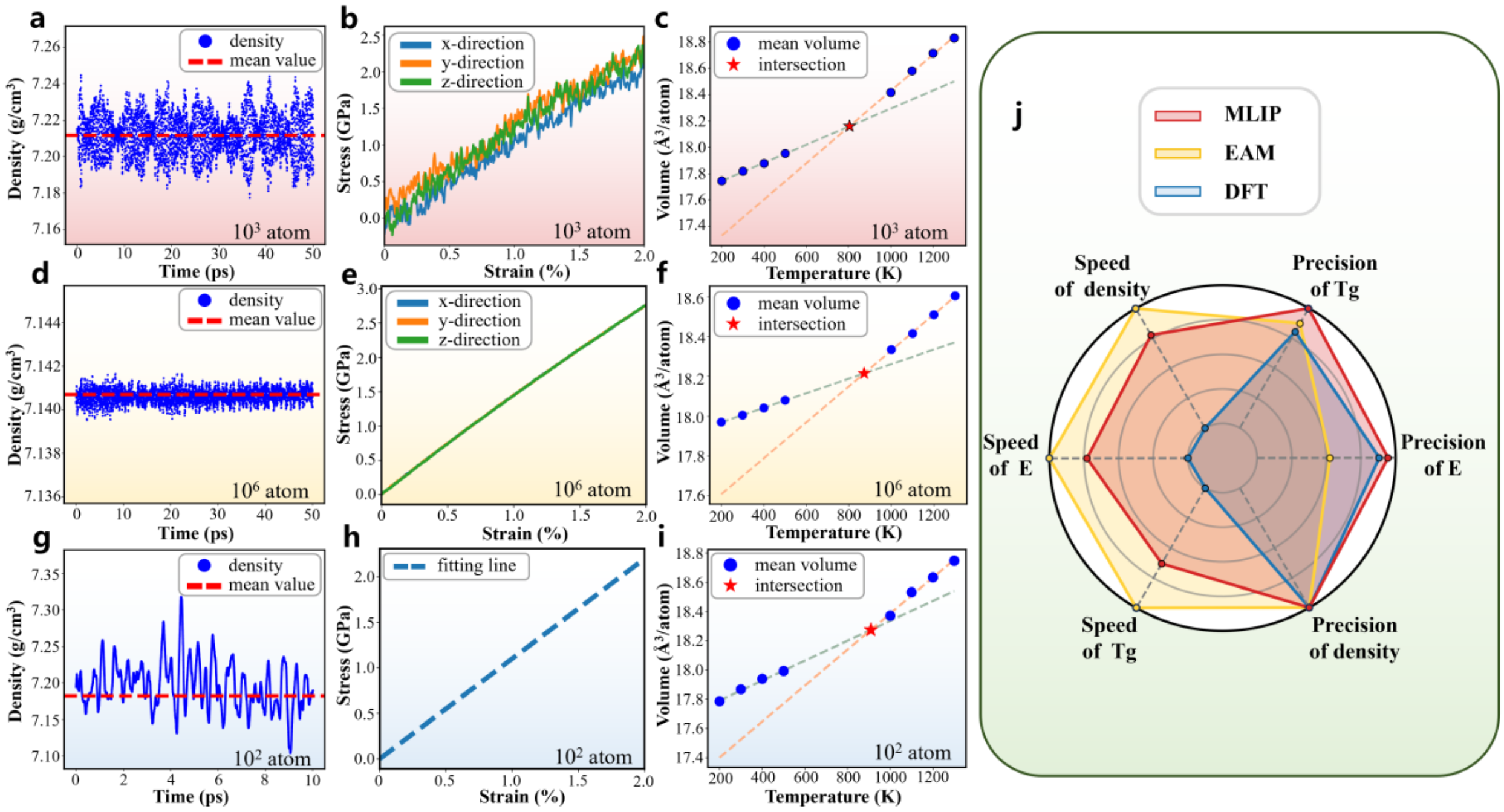

Figure 4  Comparison of different simulation methods for $Cu_{50}Zr_{50}$ metallic glass

The simulation results of density, Young's modulus and glass transition temperature are respectively calculated by (a-c) machine learning potential function, (d-f) EAM empirical potential and (g-i) first principles calculation. (j) Radar plots of accuracy and speed of three properties obtained by three methods.

To further demonstrate the advantages of using MLIPs for property simulations, we select the representative composition $Cu_{50}Zr_{50}$ and conduct simulations using three different methods: MLIPs, embedded atom method (EAM) potentials[30], and first-principles calculations. Due to differences in computational cost among the methods, the simulated systems consisted of $10^3$ atoms for MLIP, $10^6$ atoms for EAM potential, and only $10^2$ atoms for first-principles calculations. The results are presented in Figure 4. The simulations using the MLIP and EAM potential are performed using the same method. The first-principles calculations, however, employ a shorter equilibration time at each temperature to reduce computational cost and derive the Young's modulus by computing the elastic constant matrix, with the resulting stress–strain curves shown as dashed lines. Figure 4(a-c) show the results obtained using the MLIP, Figure 4(d-f) correspond to the EAM potential simulation, and Figure 4(g-i) present the results from the first-principles calculation.

It can be seen that since EAM potentials enable simulations of much larger atomic systems, which are less susceptible to thermal fluctuations, the curves in Figure 4(d) and Figure 4(e) exhibit minimal oscillation. However, empirical potentials have fewer fitting parameters, leading to lower accuracy in their simulations. In contrast, first-principles calculations, while theoretically the most accurate, are limited by their small simulation scale. This restriction makes it difficult to replicate the unique short-range order and long-range disorder of amorphous alloys, leading to certain inaccuracies in the simulation of macroscopic properties. Compared to experimental results, simulations using the MLIP exhibit the highest accuracy while maintaining a computational cost comparable to empirical potentials. As illustrated in Figure 4(j), the reciprocal of computation time required for obtaining various properties was normalized for comparative analysis. The results demonstrate that the MLIP (represented by red hexagons) exhibits superior computational efficiency, approaching the speed of empirical potentials. Furthermore, by dividing the simulation results from different methods by experimental measurements to evaluate computational accuracy, it is evident that the MLIP achieves the highest agreement with experimental values.

**2.4 Microscopic Insight into Experimental Anomalies**

During the composition development of amorphous alloys, researchers have established several influential empirical guidelines through analysis of extensive accumulated experimental data, notably including Inoue's three empirical rules[3] and the similar element substitution theory[31] for evaluating GFA. These methodologies facilitate substantial progress in amorphous alloy development. However, as explorations extend into broader compositional spaces with increasingly complex multi-element systems, emerging discrepancies between empirical predictions and experimental observations are constraining the efficiency of compositional optimization.

For instance, in the optimization of Co-Ta-B amorphous alloy properties, given that elemental Ir possesses a higher Young’s modulus than Co, the empirical rule of mixture predicts that partial substitution of Co with Ir should lead to an increased Young’s modulus. However, experimental data, as presented in Figure S5, show the opposite trend for amorphous $Co_{53-x}Ir_xTa_{10}B_{37}$ (x = 0, 10, 20, 30) samples. Figure 5(a–b) illustrates the dependence of Young's modulus and glass transition temperature on Ir content, with each plot comparing values from the rule of mixtures, experimental measurements, and simulation results. The values predicted by the rule of mixtures are referenced to the experimentally reported properties of $Co_{53}Ta_{10}B_{37}$. With increasing Ir content, the measured Young's modulus and glass transition temperature, represented by hollow triangles connected by lines, decrease continuously. This behavior is in clear contradiction to the trend predicted by the "rule of mixture", which is indicated by the blue dashed lines. In contrast, the results obtained from molecular dynamics simulations using the developed MLIP, represented by the hollow circles connected with purplish-red lines, exhibit the same decreasing trend as the experimental results.

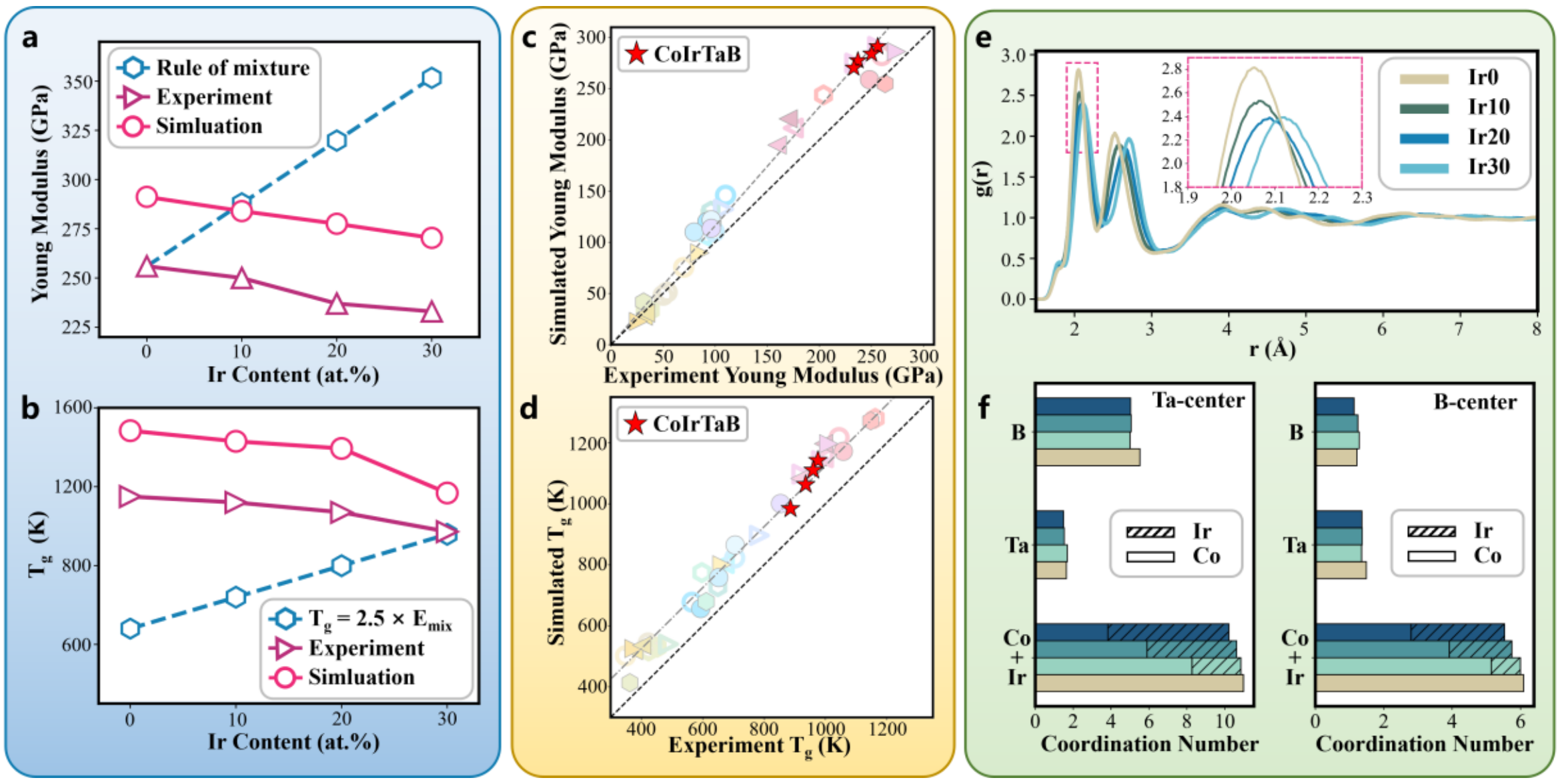

Figure 5 The application of the machine learning potential function developed in this work to the Co-Ir-Ta-B system (a-b) Comparison of empirical predictions, simulation results, and experimental values as a function of composition (c-d) Comparison between simulation results and experimental data for Co-Ir-Ta-B amorphous alloys (e-f) Structural characterization of Co-Ir-Ta-B alloys using the MLIP developed in this work

In Figures 5(c-d), the simulation values of Co-Ir-Ta-B amorphous alloys, represented by red pentagrams, still align along the dashed trend lines, exhibiting a variation trend consistent with the previously summarized pattern. Figures 5(e-f) present the structural characterization of the Co-Ir-Ta-B amorphous alloys. The radial distribution function (RDF) in Fig. 5(e) exhibits a continuous shift of the first-neighbor peak toward higher radial distances with incremental Ir addition, as highlighted in the red-dashed inset. This trend is mirrored by the systematic increase in first-neighbor peaks for predominant atomic pairs, as illustrated in Figure S6. Meanwhile, when calculating the atomic packing efficiency using the simulated structure, it can also be found that as the content of Ir elements increases, the atomic packing efficiency of the structure gradually decreases, as shown in Figure S7. These results indicate that the overall atomic structure becomes progressively more loosely packed. Furthermore, we analyze the atomic coordination environment around each element, as shown in Figure 5(f). It can be seen that the addition of Ir does not significantly alter the coordination environment of the Co-Ta-B amorphous alloy. This observation supports the theory of similar element substitution. Ir, being in the same group as Co, induces minimal changes in the local atomic structure upon substitution and therefore does not noticeably deteriorate the GFA of the alloy.

The consistency between experiment and MLIP simulations results indicates that the potential constructed in this work can accurately capture the compositional dependence of the material properties. The design approach based on MLIP offers an efficient tool for accelerating the development of property-targeted compositions in amorphous alloys. This approach captures material property variations that fall outside the predictive scope of traditional empirical rules, contributing valuable insights to the advancement of amorphous alloy design. Furthermore, considering that Ir is a precious metal, employing the developed potential for compositional design not only accelerates materials development but also reduces experimental costs.

## 3 Conclusion

In this work, we establish a workflow for MLIP development and property prediction of amorphous alloys, based on data derived from first-principles calculations. A DFT dataset covering 19 representative binary and ternary amorphous alloy systems is constructed. Starting from the pretrained CHGNet model, we train a MLIP capable of achieving near-DFT accuracy across all 19 systems. This MLIP enables structural modeling as well as predictions of key macroscopic properties such as density, Young's modulus, and glass transition temperature. The simulation results exhibit strong agreement with experimentally reported trends. With appropriate systematic corrections, it can provide valuable guidance for compositional design of amorphous alloys. Furthermore, we extend the application of the trained potential to amorphous alloy systems beyond the training set, and find that it still accurately reproduces the macroscopic properties of these more complex compositions. Compared to conventional empirical potentials and first-principles calculations, the proposed MLIP offers a compelling balance of high efficiency and high accuracy. In addition to property prediction, it also provides access to atomistic configurations, supporting microscopic interpretation and mechanistic insight during composition development in experiment.

Amorphous alloys, as materials with unique structures characterized by short-range order and long-range disorder, inherently require larger spatial scales for accurate simulation. This makes them particularly suitable for computational simulations using MLIPs. Given the dual advantages of accuracy and computational efficiency demonstrated by the MLIP developed in this work, we highlight this MLIP for high-throughput compositional screening using this MLIP to optimize 19 amorphous alloy systems comprising 25 elements included in the training dataset. Based on the theory of similar element substitution, numerous hypothetical new amorphous alloy compositions can be generated and investigated through molecular dynamics simulations using the MLIP in this work. This enables high-accuracy composition-property mapping, laying the foundation for data-driven design of new amorphous alloys. Moreover, for studies aimed at mechanistic understanding of specific systems, the MLIP provides access to atomic configurations of amorphous alloys, facilitating the investigation of microscopic phenomena and underlying mechanisms.

## 4 Method

### 4.1 DFT calculation

The first-principles calculations are carried out using the Vienna Ab Initio Simulation Package (VASP) with the projector augmented wave (PAW) method. The exchange-correlation functional is approximated using the generalized gradient approximation (GGA) with the Perdew-Burke-Ernzerhof (PBE) functional. Each structure in the training set and test set contains 100 atoms. During the AIMD process, we use a cut-off energy set to 400 eV and a 1×1×1 Γ-centered mesh. For static calculations, the plane-wave cut-off energy is set to 520 eV and the k-mesh grid is 2×2×2. The energy and force convergence criteria are set to $10^{-5}$ eV and 0.01 eV/Å, respectively.

### 4.2 CHGNet training

Based on the CHGNet pre-training model, the energies, forces and stresses in the data set are read for training. All data is divided into training set, validation set and test set according to 8:1:1. The model is trained for 500 iterations with an initial learning rate of $1\times10^{-3}$. The mean squared error (MSE) is used as the evaluation metric for the loss function, and the model with the lowest validation error is selected for testing on an independent dataset.

### 4.3 Molecular dynamics simulation

The best-trained model is employed as the potential function to carry out molecular dynamics simulations using the Large-scale Atomic/Molecular Massively Parallel Simulator (LAMMPS). Molecular dynamics simulations are carried out on systems composed of 1000 atoms. The rapid quenching simulation procedure consists of a 10 ps melting stage at 2500 K, a 22 ps rapid cooling stage, and a final 10 ps relaxation stage at 300 K, all conducted under NPT conditions. Uniaxial tensile simulations are performed by first equilibrating the amorphous structure for 10 ps, followed by the application of uniaxial strain along the X, Y, or Z direction until a total deformation of 2% is reached.

### 4.4 Experiment Details

For Co-Ir-Ta-B system, amorphous alloy rods with a diameter of 1 mm are fabricated via copper mold casting, using master alloys pre-melted by high-frequency induction heating and vacuum arc melting. The amorphous structure is identified through X-ray diffraction. The density is measured using the Archimedes' method, while the Young's modulus is determined by the ultrasonic resonance technique. The glass transition temperature is evaluated using differential scanning calorimetry (DSC).

## Acknowledgements

This study was funded by the National Natural Science Foundation of China (Grants Nos. 52473227, 52171150, 52172258) and the Strategic Priority Research Program of Chinese Academy of Sciences (Grant Nos. XDB0500200, XDB1040300). We acknowledge both the National Supercomputer Center in Tianjin and the ORISE Supercomputer for providing computational resources.

## Author information

Authors and Affiliations

School of Materials Science and Engineering, Key Laboratory of Aerospace Materials and Performance (Ministry of Education), Beihang University, Beijing, China

Xuhe Gong, Hengbo Zhao, Ran Li, Tao Zhang

Institute of Physics, Chinese Academy of Sciences, Beijing, China

Xuhe Gong, Xiao Fu, Jingchen Lian, Qifan Yang, Ruijuan Xiao, Hong Li

Contributions

R.L. and R.X. designed and guided the completion of the method. X.G. and R.X. constructed the machine learning interatomic potential, X.G. performed all simulations. X.F., J.L. and Q.Y. provided suggestions for the theoretical calculations. X.G. and H.Z. designed and conducted the experiment. All the authors participated in the analysis of the data and discussions of the results, as well as in preparing the paper.

Corresponding authors

Correspondence to Ran Li, Ruijuan Xiao

## Competing interests

All authors declare no financial or non-financial competing interests.

## Data availability

The data supporting this study's findings are available within this article and its Supplementary Information. Additional data are available from the corresponding authors on reasonable request. All the data and codes for machine learning interatomic potential can be accessed via https://github.com/xuhegg/MG_MLIP2025. All dataset used in this study will be publicly released after the official publication of the article.

## Reference

1. Turnbull, D. Under what conditions can a glass be formed? *Contemp. Phys.* **10**, 473–488 (1969).
2. Greer, A. L. Confusion by design. *Nature* **366**, 303–304 (1993).
3. Inoue, A. Stabilization of metallic supercooled liquid and bulk amorphous alloys. *Acta Mater.* **48**, 279–306 (2000).
4. Shi, L. & Yao, K. Composition design for Fe-based soft magnetic amorphous and nanocrystalline alloys with high Fe content. *Mater. Des.* **189**, 108511 (2020).
5. Li, M.-X. *et al.* High-temperature bulk metallic glasses developed by combinatorial methods. *Nature* **569**, 99–103 (2019).
6. Li, H. F. *et al.* In vitro and in vivo studies on biodegradable CaMgZnSrYb high-entropy bulk metallic glass. *Acta Biomater.* **9**, 8561–8573 (2013).
7. Hart, G. L. W., Mueller, T., Toher, C. & Curtarolo, S. Machine learning for alloys. *Nat. Rev. Mater.* **6**, 730–755 (2021).
8. Butler, K. T., Davies, D. W., Cartwright, H., Isayev, O. & Walsh, A. Machine learning for molecular and materials science. *Nature* **559**, 547–555 (2018).
9. Gao, C. *et al.* Innovative materials science via machine learning. *Adv. Funct. Mater.* **32**, 2108044 (2022).
10. Bourel, M. *et al.* Machine learning methods for imbalanced data set for prediction of faecal contamination in beach waters. *Water Res.* **202**, 117450 (2021).
11. Han, X. *et al.* In situ observation of structural evolution and phase engineering of amorphous materials during crystal nucleation. *Adv. Mater.* **34**, 2206994 (2022).
12. Jia, Z. *et al.* A self-supported high-entropy metallic glass with a nanosponge architecture for efficient hydrogen evolution under alkaline and acidic conditions. *Adv. Funct. Mater.* **31**, 2101586 (2021).
13. Sivonxay, E. & Persson, K. A. Density functional theory assessment of the lithiation thermodynamics and phase evolution in Si-based amorphous binary alloys. *Energy Storage Mater.* **53**, 42–50 (2022).
14. Lee, M., Lee, C.-M., Lee, K.-R., Ma, E. & Lee, J.-C. Networked interpenetrating connections of icosahedra: Effects on shear transformations in metallic glass. *Acta Mater.* **59**, 159–170 (2011).
15. Zhang, L. *et al.* Heterogeneity of microstructures in a Cu–Zr based amorphous alloy composite reinforced by crystalline phases. *Compos. Part B Eng.* **262**, 110823 (2023).
16. Iabbaden, D., Amodeo, J., Fusco, C., Garrelie, F. & Colombier, J.-P. Dynamics of Cu–Zr metallic glass devitrification under ultrafast laser excitation revealed by atomistic modeling. *Acta*

*Mater.* **263**, 119487 (2024).
17. Kulichenko, M. *et al.* Data generation for machine learning interatomic potentials and beyond. *Chem. Rev.* **124**, 13681–13714 (2024).
18. Yang, Q. *et al.* Atomic insight into $Li^+$ ion transport in amorphous electrolytes $Li_xAlO_yCi_{3+x-2y}$ ($0.5 \leq x \leq 1.5$, $0.25 \leq y \leq 0.75$). *J. Mater. Chem. A* **13**, 2309–2315 (2025).
19. Wan, K., He, J. & Shi, X. Construction of High Accuracy Machine Learning Interatomic Potential for Surface/Interface of Nanomaterials—A Review. *Adv. Mater.* **36**, 2305758 (2024).
20. Deng, B. *et al.* CHGNet as a pretrained universal neural network potential for charge-informed atomistic modelling. *Nat. Mach. Intell.* **5**, 1031–1041 (2023).
21. Zhang, D. *et al.* DPA-2: A large atomic model as a multi-task learner. *Npj Comput. Mater.* **10**, 293 (2024).
22. Liang, T. *et al.* NEP89: Universal neuroevolution potential for inorganic and organic materials across 89 elements. *arXiv.org* https://arxiv.org/abs/2504.21286v2.
23. Li, F. *et al.* Oxidation-induced superelasticity in metallic glass nanotubes. *Nat. Mater.* **23**, 52–57 (2024).
24. Zhao, R. *et al.* Development of a neuroevolution machine learning potential of pd-cu-ni-P alloys. *Mater. Des.* **231**, 112012 (2023).
25. Thompson, A. P. *et al.* LAMMPS - a flexible simulation tool for particle-based materials modeling at the atomic, meso, and continuum scales. *Comput. Phys. Commun.* **271**, 108171 (2022).
26. Kresse, G. & Furthmüller, J. Efficient iterative schemes for *ab initio* total-energy calculations using a plane-wave basis set. *Phys. Rev. B* **54**, 11169–11186 (1996).
27. Jain, A. *et al.* Commentary: The materials project: a materials genome approach to accelerating materials innovation. *APL Mater.* **1**, 011002 (2013).
28. Cheng, Y. Q. & Ma, E. Atomic-level structure and structure–property relationship in metallic glasses. *Prog. Mater. Sci.* **56**, 379–473 (2011).
29. Debenedetti, P. G. & Stillinger, F. H. Supercooled liquids and the glass transition. *Nature* **410**, 259–267 (2001).
30. Mendelev, M. I., Sordelet, D. J. & Kramer, M. J. Using atomistic computer simulations to analyze x-ray diffraction data from metallic glasses. *J. Appl. Phys.* **102**, 043501 (2007).
31. Li, R., Pang, S., Ma, C. & Zhang, T. Influence of similar atom substitution on glass formation in (La–Ce)–Al–Co bulk metallic glasses. *Acta Mater.* **55**, 3719–3726 (2007).

**Accelerating Amorphous Alloy Discovery: Data-Driven Property Prediction via General-Purpose Machine Learning Interatomic Potential**

Xuhe Gong [a,b], Hengbo Zhao [a], Xiao Fu [b], Jingchen Lian [b], Qifan Yang [b],

Ran Li [a,*], Ruijuan Xiao [b,*], Tao Zhang [a], Hong Li [b]

[a] *School of Materials Science and Engineering, Key Laboratory of Aerospace Materials and Performance (Ministry of Education), Beihang University, Beijing 100191, China;*

[b] *Institute of Physics, Chinese Academy of Sciences, Beijing 100190, China*

*E-mail: liran@buaa.edu.cn (R. Li), rjxiao@iphy.ac.cn (R. Xiao).

## Supplementary Information

### Table of Contents

Table S1 Comparison between the simulated physical property data of amorphous alloy compositions included in the training set and corresponding experimentally reported values

| Composition | Exp_ρ (g/cm$^3$) | Sim_ρ (g/cm$^3$) | Exp_E (GPa) | Sim_E (GPa) | Exp_$T_g$ (K) | Sim_$T_g$ (K) | Ref. |
|---|---|---|---|---|---|---|---|
| $Au_{55}Cu_{25}Si_{20}$ | 12.20 | 12.893 | 69.8 | 75.41 | 348 | 502 | [1] |
| $Mg_{58.5}Cu_{30.5}Y_{11}$ | 3.547 | 3.496 | 53.9 | 51.31 | 422 | 544 | [2] |
| $Mg_{65}Cu_{25}Y_{10}$ | 3.28 | 3.175 | 50.1 | 48.15 | 425 | 523 | [3] |
| $La_{60}Al_{15}Co_{25}$ | 6.222 | 6.085 | 37.7 | 33.54 | 451 | 530 | [4] |
| $La_{70}Al_{10}Co_{20}$ | 6.218 | 6.167 | 32.7 | 32.89 | 423 | 518 | [4] |
| $Pt_{60}Ni_{15}P_{25}$ | 15.70 | 14.625 | 96.1 | 123.38 | 488 | 540 | [5] |
| $Pd_{40}Ni_{40}P_{20}$ | 9.405 | 9.013 | 108 | 145.95 | 576 | 678 | [6] |
| $Ti_{45}Zr_{20}Be_{35}$ | 4.59 | 4.564 | 96.8 | 119.45 | 597 | 775 | [7] |
| $Pd_{81}Si_{19}$ | 10.61 | 9.886 | 96 | 130.37 | 649 | 729 | [8] |
| $Cu_{50}Zr_{50}$ | 7.404 | 7.212 | 87 | 105.70 | 670 | 805 | [3] |
| $Cu_{45}Zr_{45}Al_{10}$ | 7.204 | 6.807 | 99.1 | 110.33 | 704 | 823 | [3,9] |
| $Cu_{66}Hf_{34}$ | / | 10.869 | 108 | 132.26 | 784 | 897 | [10] |
| $Co_{53}Ta_{10}B_{37}$ | 9.291 | 9.352 | 256 | 291.23 | 975 | 1142 | [11] |
| $Co_{61}Ta_{6}B_{33}$ | 8.814 | 8.915 | 234 | 274.97 | 923 | 1098 | [11] |
| $Ni_{61}Ta_{39}$ | / | 12.777 | 175 | 211.49 | 993 | 1152 | [12] |
| $Mo_{51}Co_{34}B_{14}$ | / | 9.274 | 260 | 282.20 | 1044 | 1218 | [13] |
| $Ir_{35}Ni_{25}Ta_{40}$ | / | 16.570 | 204 | 243.86 | 1162 | 1281 | [14] |

Table S2 Comparison between the simulated physical property data of extrapolated composition and corresponding experimentally reported values

| Composition | Exp_ρ (g/cm$^3$) | Sim_ρ (g/cm$^3$) | Exp_E (GPa) | Sim_E (GPa) | Exp_$T_g$ (K) | Sim_$T_g$ (K) | Ref. |
|---|---|---|---|---|---|---|---|
| $La_{34}Ce_{34}Al_{10}Cu_{20}Co_{2}$ | 6.492 | 6.932 | 31.27 | 41.77 | 362 | 414 | [15] |
| $Ca_{55}Mg_{18}Zn_{11}Cu_{16}$ | 2.411 | 2.410 | 31.0 | 27.91 | 373 | 520 | [16] |
| $Ca_{65}Mg_{15}Zn_{20}$ | 2.050 | 2.107 | 26.4 | 22.63 | 375 | 526 | [16] |
| $Ca_{50}Mg_{20}Cu_{30}$ | 2.589 | 2.596 | 33.2 | 30.88 | 401 | 537 | [16] |
| $Pd_{40}Cu_{30}Ni_{10}P_{20}$ | 9.28 | 8.841 | 92 | 120.26 | 593 | 657 | [3] |
| $Ti_{32.8}Zr_{30.2}Ni_{5.3}Cu_{9}Be_{22.7}$ | 5.541 | 5.457 | 97.8 | 120.28 | 611 | 680 | [17] |
| $Sc_{36}Al_{24}Co_{20}Y_{20}$ | 4.214 | 4.166 | 85.2 | 89.66 | 662 | 801 | [18] |
| $Zr_{62}Cu_{15.5}Ni_{12.5}Al_{10}$ | 6.615 | 6.567 | 79.65 | 109.99 | 652 | 759 | [19] |
| $Zr_{46}Cu_{30.1}Ag_{8.4}Al_{8}Be_{7.5}$ | 6.933 | 6.783 | 96.3 | 121.93 | 705 | 866 | [20] |
| $Cr_{15}Fe_{41}Co_{7}Mo_{14}C_{15}B_{6}Y_{2}$ | 7.867 | 7.701 | 214 | 244.19 | 853 | 1002 | [21] |
| $Ni_{52}Co_{10}Nb_{33}Ta_{5}$ | / | 9.155 | 159.2 | 195.00 | 916 | 1084 | [22] |
| $Ta_{42}Ni_{40}Co_{18}$ | 12.98 | 13.044 | 170 | 220.57 | 993 | 1180 | [23] |
| $Co_{27}Os_{27}Ta_{10}B_{36}$ | 14.062 | 13.655 | 274 | 283.32 | 1010 | 1198 | [24] |
| $Co_{26}Cr_{26}Mo_{26}Nb_{7}B_{15}$ | 8.41 | 8.072 | 248 | 258.94 | 1058 | 1173 | [25] |
| $Ir_{35}Ni_{20}Ta_{40}B_{5}$ | / | 16.556 | 263 | 254.93 | 1147 | 1273 | [14] |

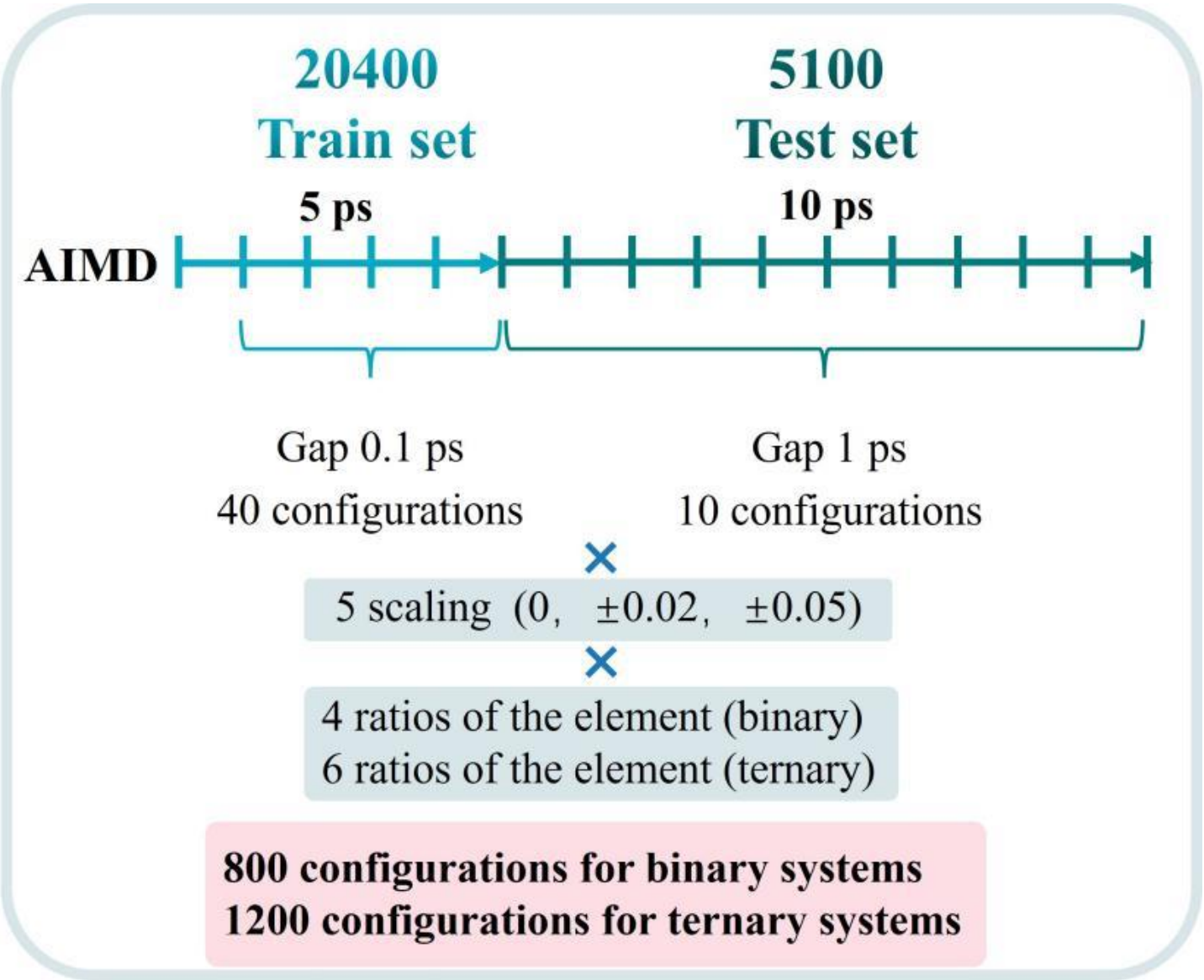

**Figure S1 The specific construction method of the dataset**

The structures in dataset are extracted from ab initio molecular dynamics (AIMD) trajectories at fixed intervals. Specifically, from a 15 ps AIMD simulation, configurations between 2 ps and 5 ps are sampled every 0.1 ps to construct the training set, while those between 5 ps and 15 ps are sampled every 1 ps to form the test set. For both the training and test sets, each configuration is further augmented by applying four uniform scaling factors (0.95, 0.98, 1.02, and 1.05) to the cell dimensions, relative to the room-temperature equilibrium volume. To ensure compositional diversity across each binary or ternary system, compositions are selected at regular intervals based on atomic ratios. For a binary alloy system A–B (where A and B represent distinct elements), four compositions are chosen: $A_{20}B_{80}$, $A_{40}B_{60}$, $A_{60}B_{40}$, and $A_{80}B_{20}$. For ternary systems A–B–C (where A, B and C represent distinct elements), six compositions are selected: $A_{20}B_{20}C_{60}$, $A_{20}B_{40}C_{40}$, $A_{20}B_{60}C_{20}$, $A_{40}B_{20}C_{40}$, $A_{40}B_{40}C_{20}$, and $A_{60}B_{20}C_{20}$.

In total, each binary system contributes 800 configurations to the training set and 200 configurations to the test set, while the training and test sets for each ternary system consist of 1200 and 300 configurations, respectively.

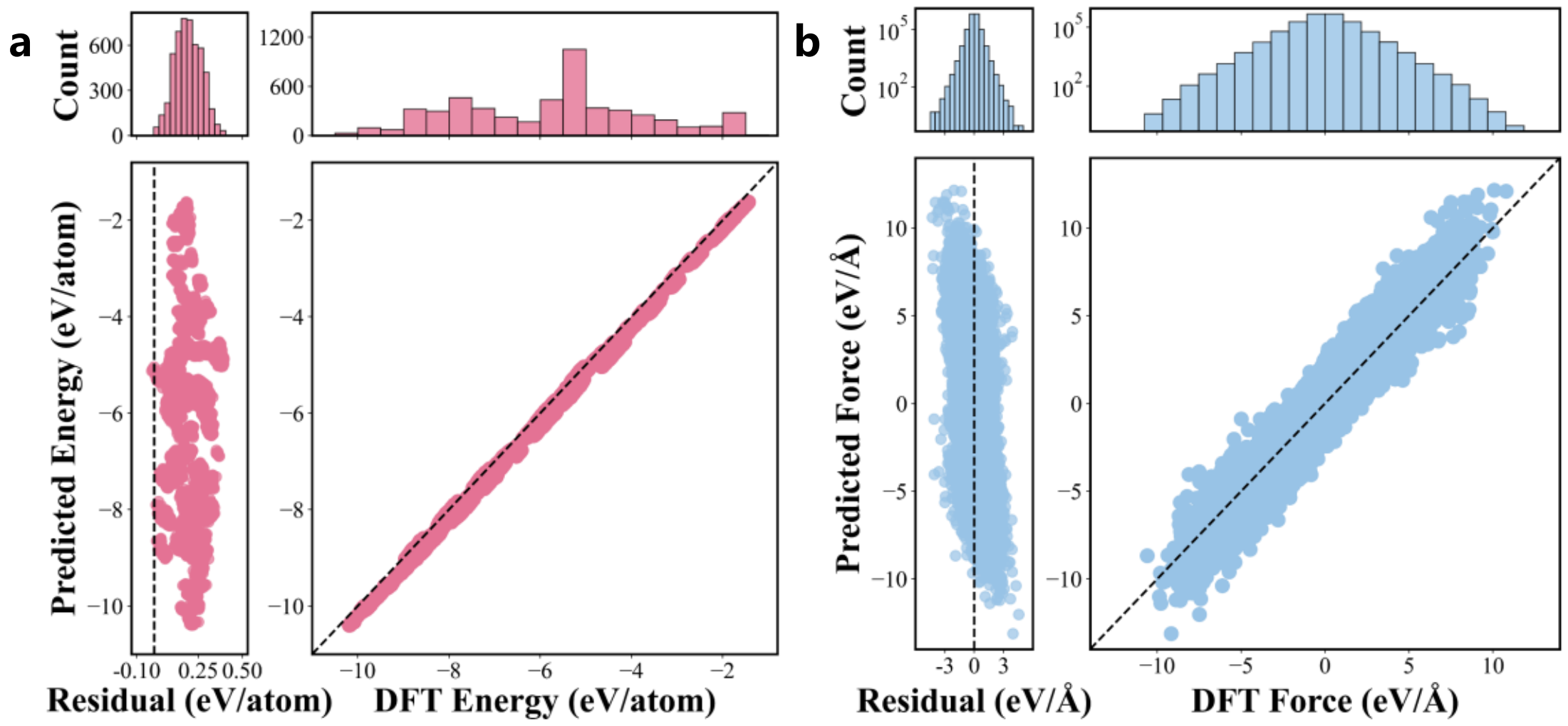

**Figure S2 The prediction results and error statistics of the pretrained model on the test set (a) energies (b) forces**

The pretrained CHGNet potential is trained on data from the Materials Project (MP) database, where disordered and high-energy structures are comparatively scarce. As a result, it exhibits relatively large prediction errors for such configurations. On the independent test set used in this work, the pretrained model yields a mean absolute error (MAE) of 191.555 meV/atom for energy and 275.163 meV/Å for forces. Therefore, fine-tuning the model is essential, as it significantly improves the accuracy of molecular dynamics simulations for amorphous structures.

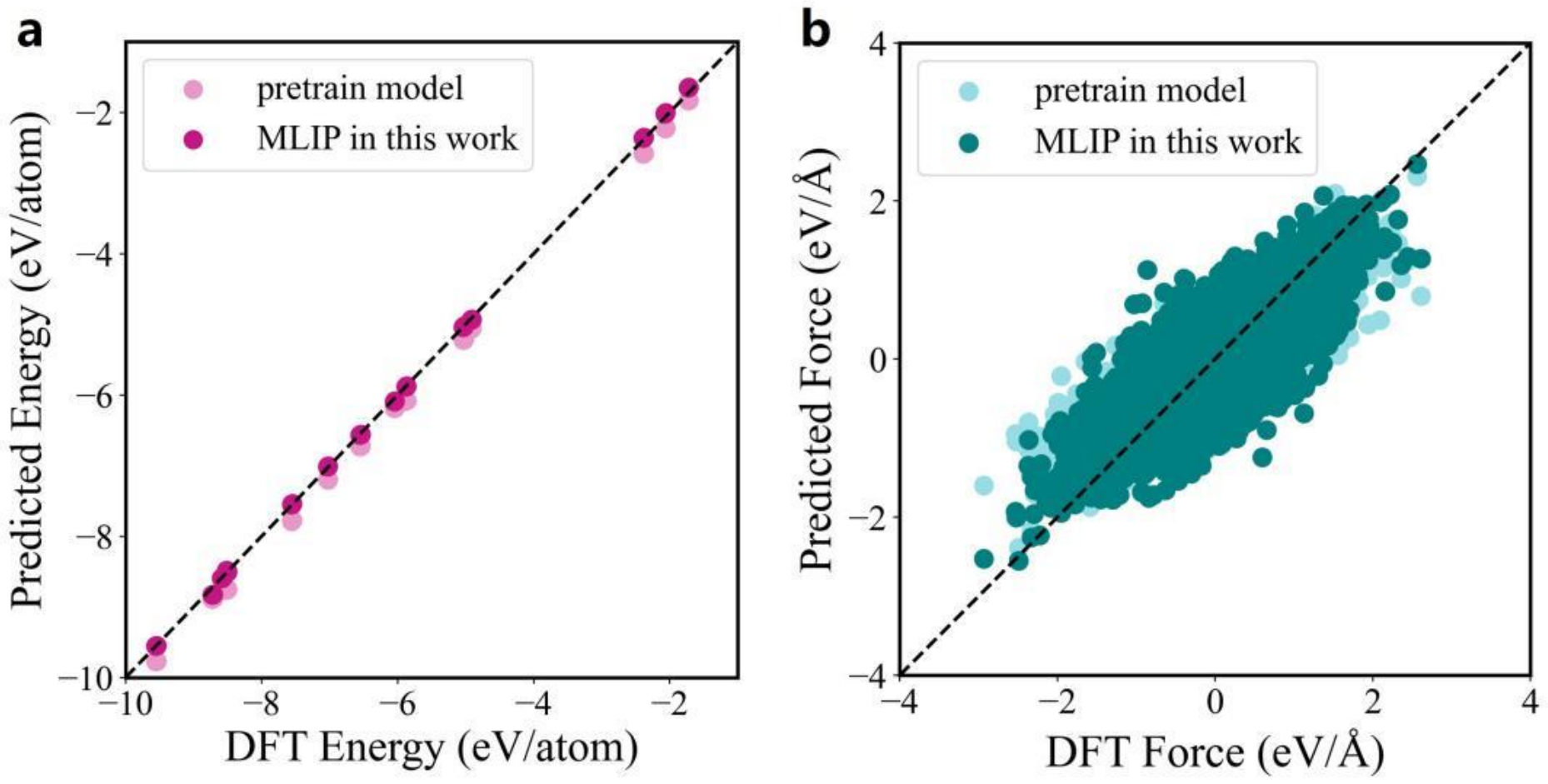

**Figure S3 Performance comparison between the pretrained model and our developed model on the extrapolated compositional test set.**

(a) Energy prediction performance; (b) Force prediction performance

Using the MLIP developed in this work, we simulate the rapid quenching process of the extrapolated compositions outside the training set. Static energy calculations are conducted by extracting snapshots every 1 ps from a 10 ps NPT simulation at 300 K, resulting in an additional independent test set consisting of 150 configurations across 15 alloy compositions. Although some atomic species or interactions in these extrapolated structures appear only in the pretrained dataset, the fine-tuned model significantly improves energy prediction accuracy. As shown in Supplementary Figure S3, the fine-tuned model reduces the MAE to 27.216 meV/atom for energy and 145.455 meV/Å for forces, compared to 183.039 meV/atom and 166.188 meV/Å with the pretrained model. The extrapolated compositions are derived from training systems via substitution with similar elements, resulting in local atomic environments that closely resemble those the model has learned. This structural similarity enhances the model's ability to generalize and predict properties for previously unseen compositions.

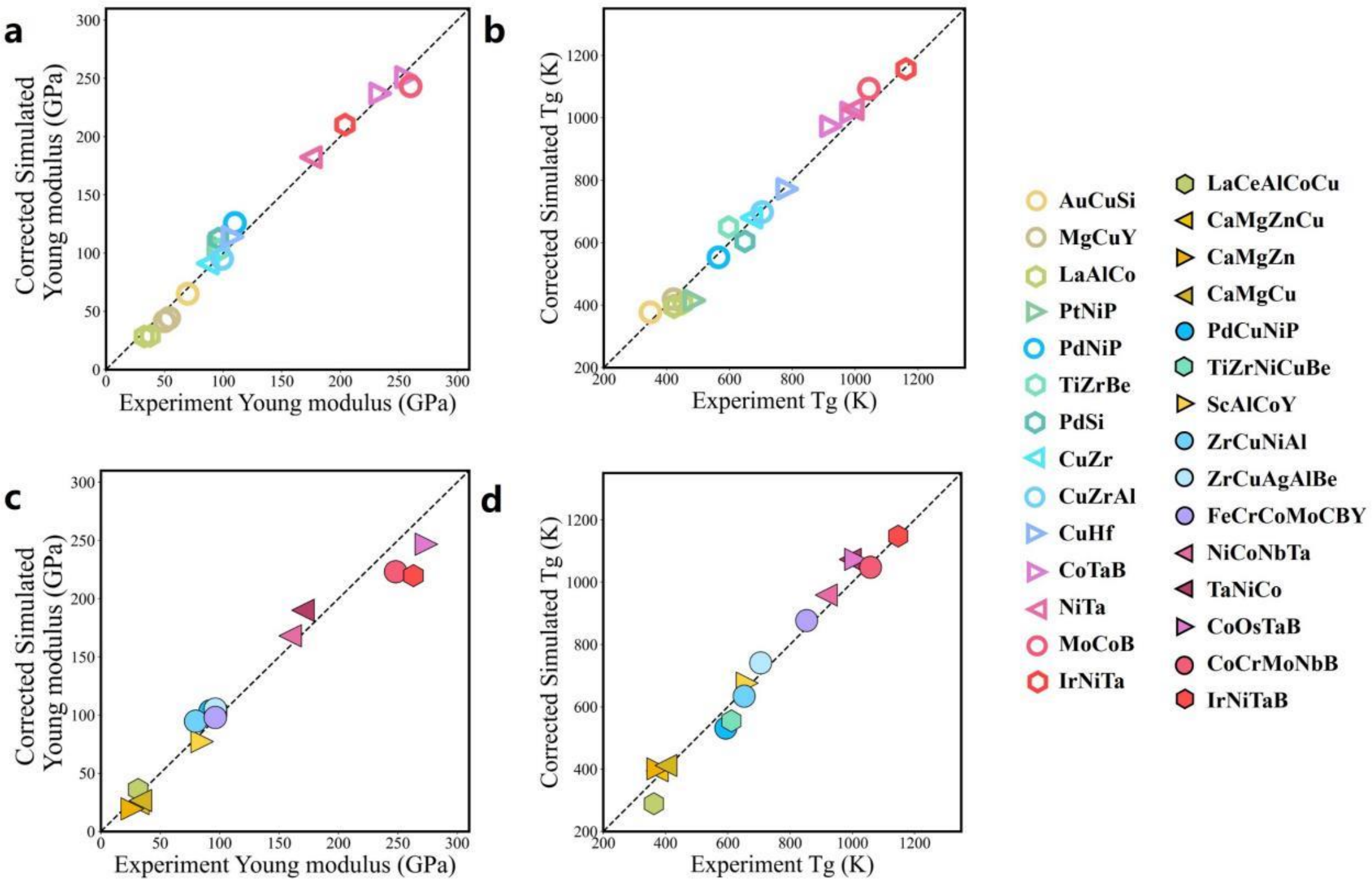

**Figure S4 Corrected predictions of macroscopic properties for amorphous alloys based on MLIP simulations**

(a-b) Comparison between corrected simulation results and experimental data for compositions within the training set: (a) Young's modulus; (b) Glass transition temperature.

(c-d) Comparison between corrected simulation results and experimental data for extrapolated compositions outside the training set: (c) Young's modulus; (d) Glass transition temperature.

Following the application of a uniform correction, the simulated Young's modulus and glass transition temperatures exhibit strong agreement with experimental values. As depicted in Figure S4(a), the corrected Young's modulus for systems within the training set demonstrates an MAE of 9.36 GPa relative to experimental data, while Figure S3(b) shows that the corrected glass transition temperature yields an MAE of 32 K. Notably, for compositions extrapolated beyond the training set, the corrected Young's modulus retains good predictive capability with an MAE of 15.85 GPa as shown in Figure S3(c), and the corrected glass transition temperature maintains accuracy with an MAE of 35 K illustrated in Figure S3(d).

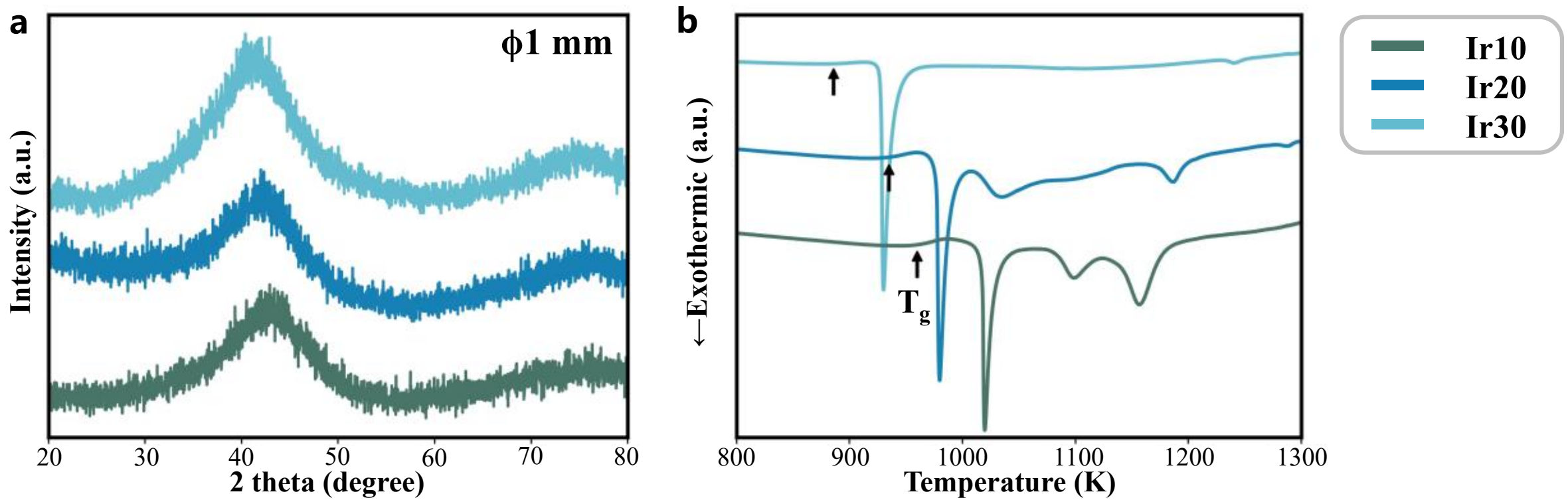

**Figure S5 $Co_{53-x}Ir_xTa_{10}B_{37}$ (x = 10, 20, 30) experiment results**

(a) X-ray diffraction pattern (b) differential scanning calorimetric curve

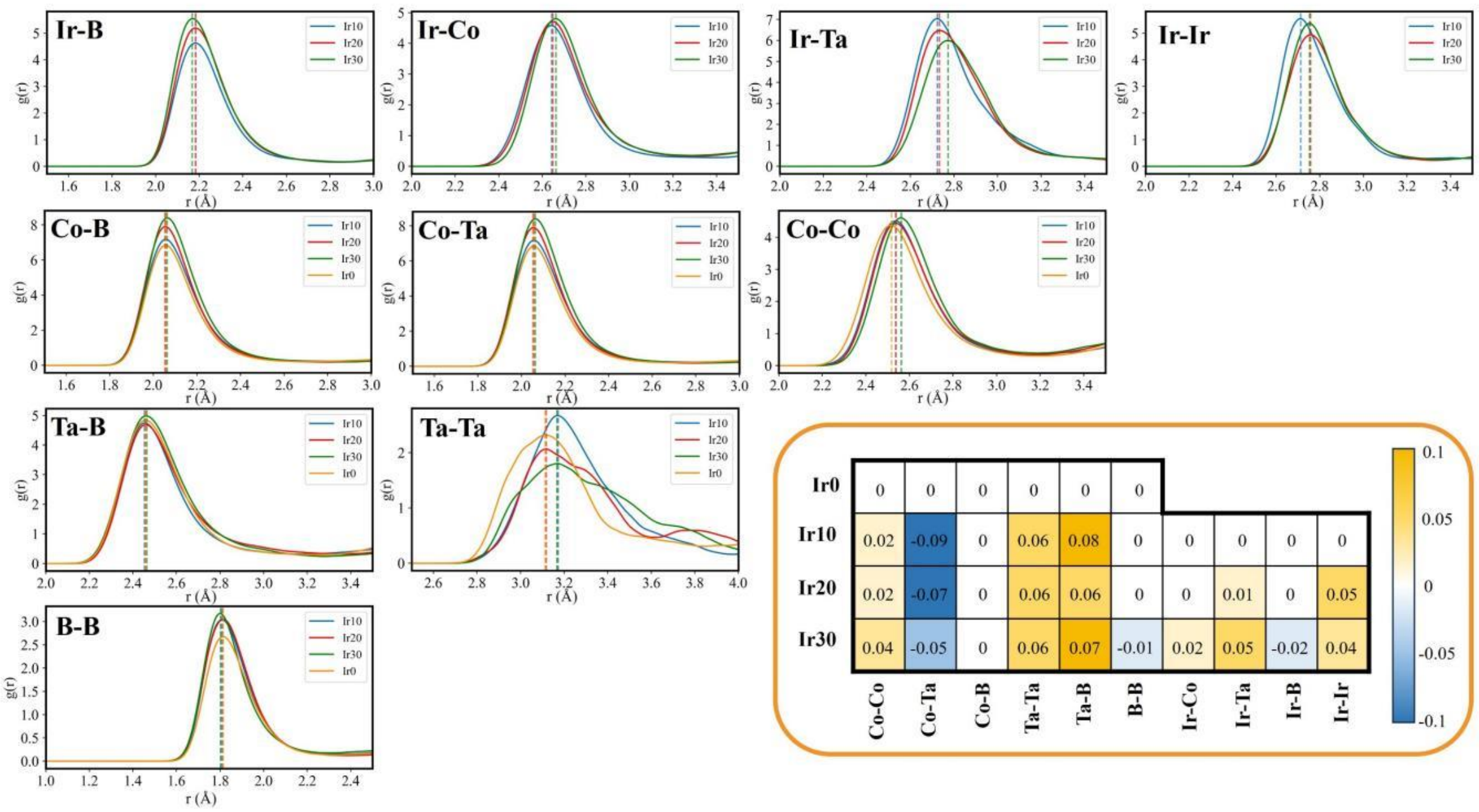

**Figure S6 Pair distribution function of $Co_{53-x}Ir_xTa_{10}B_{37}$ (x = 0, 10, 20, 30)**

A statistical analysis is conducted on the pair distribution functions (PDFs) of the $Co_{53-x}Ir_xTa_{10}B_{37}$ alloy series (x = 0, 10, 20, 30). The peak position shifts of the first PDF peaks for each atomic pair, relative to those in $Co_{53}Ta_{10}B_{37}$ are shown in the bottom-right panel (with $Co_{43}Ir_{10}Ta_{10}B_{37}$ used as the reference for Ir-centered pairs). Positive and negative shifts are indicated in orange and blue, respectively. Except for the Co–Ta pair, most atomic pair distances increase progressively with higher Ir content, suggesting a more loosely packed atomic arrangement.

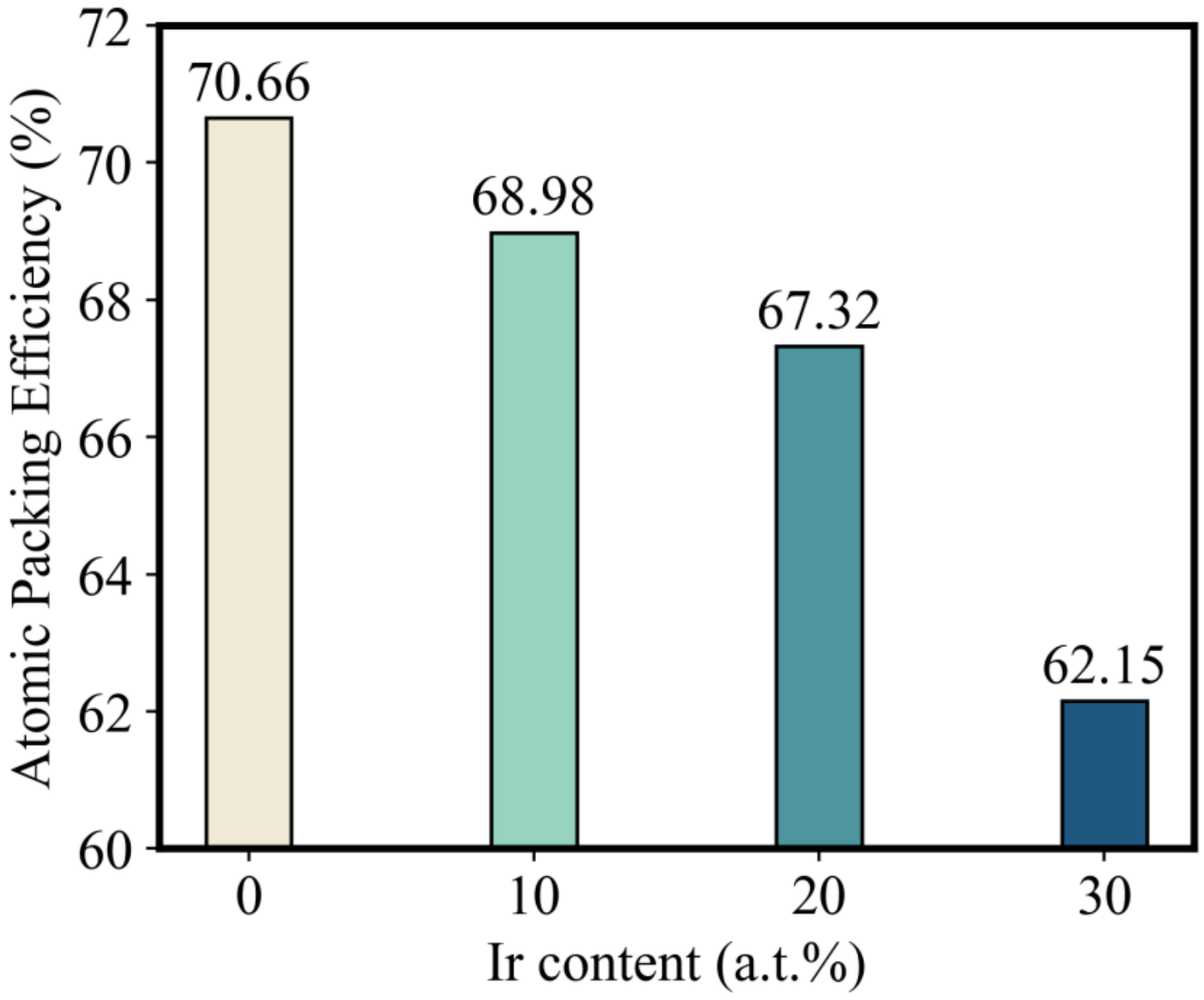

**Figure S7 Atomic packing efficiency of $Co_{53-x}Ir_xTa_{10}B_{37}$ (x = 0, 10, 20, 30)**

The atomic packing efficiency (APE) serves as an indicator of the degree of atomic arrangement tightness within a material. The APE is calculated using the following formula:

$$APE = \frac{\sum n_i \frac{4}{3}\pi {r_i}^3}{V}$$

where $n_i$ represents the number of atoms of element i in the structure, $r_i$ denotes the atomic radius of element i, and V is the volume of the structure. We calculate the APE of the four compositions at 300 K and observe a gradual decrease with increasing Ir content. This trend reflects a progressive loosening of the atomic structure, which is consistent with the observed decline in Young's modulus.

## Reference

1. Schroers, J., Lohwongwatana, B., Johnson, W. L. & Peker, A. Gold based bulk metallic glass. *Applied Physics Letters* **87**, 061912 (2005).
2. Zheng, Q., Ma, H., Ma, E. & Xu, J. Mg–Cu–(Y, Nd) pseudo-ternary bulk metallic glasses: The effects of Nd on glass-forming ability and plasticity. *Scripta Materialia* **55**, 541–544 (2006).
3. Wang, W. H. The elastic properties, elastic models and elastic perspectives of metallic glasses. *Progress in Materials Science* **57**, 487–656 (2012).
4. Li, R., Pang, S., Ma, C. & Zhang, T. Influence of similar atom substitution on glass formation in (la–ce)–al–co bulk metallic glasses. *Acta Materialia* **55**, 3719–3726 (2007).
5. Chen, H. S., Krause, J. T. & Sigety, E. A. Thermal expansion and density of glassy Pd-Ni-P and Pd-Ni-P alloys. *Journal of Non-Crystalline Solids* **13**, 321–327 (1974).
6. Lambson, E. F. *et al.* Elastic behavior and vibrational anharmonicity of a bulk Pd40Ni40P20 metallic glass. *Phys. Rev. B* **33**, 2380–2385 (1986).
7. Duan, G., Wiest, A., Lind, M. L., Kahl, A. & Johnson, W. L. Lightweight Ti-based bulk metallic glasses excluding late transition metals. *Scripta Materialia* **58**, 465–468 (2008).
8. Yao, K. & Chen, N. Pd-Si binary bulk metallic glass. *Sci. China Ser. G-Phys. Mech. As* **51**, 414–420 (2008).
9. Cheung, T. L. & Shek, C. H. Thermal and mechanical properties of Cu–Zr–Al bulk metallic glasses. *Journal of Alloys and Compounds* **434–435**, 71–74 (2007).
10. Duan, G., Xu, D. & Johnson, W. L. High copper content bulk glass formation in bimetallic Cu-Hf system. *Metall Mater Trans A* **36**, 455–458 (2005).
11. Wang, J. *et al.* Ultra-high strength Co–Ta–B bulk metallic glasses: Glass formation, thermal stability and crystallization. *Journal of Alloys and Compounds* **860**, 158398 (2021).
12. Wang, Y., Wang, Q., Zhao, J. & Dong, C. Ni–Ta binary bulk metallic glasses. *Scripta Materialia* **63**, 178–180 (2010).
13. Lai, L. *et al.* High-temperature Mo-based bulk metallic glasses. *Scripta Materialia* **203**, 114095 (2021).
14. Li, M.-X. *et al.* High-temperature bulk metallic glasses developed by combinatorial methods. *Nature* **569**, 99–103 (2019).
15. Liu, X. F., Wang, R. J., Zhao, D. Q., Pan, M. X. & Wang, W. H. Bulk metallic glasses based on binary cerium and lanthanum elements. *Applied Physics Letters* **91**, 041901 (2007).
16. Senkov, O. N., Miracle, D. B., Keppens, V. & Liaw, P. K. Development and characterization of low-density ca-based bulk metallic glasses: An overview. *Metall Mater Trans A* **39**, 1888–1900 (2008).
17. Tang, M. Q. *et al.* TiZr-base bulk metallic glass with over 50 mm in diameter. *Journal of Materials Science & Technology* **26**, 481–486 (2010).
18. Xi, X. K. *et al.* Bulk scandium-based metallic glasses. *J. Mater. Res.* **20**, 2243–2247 (2005).
19. Liu, Y. H. *et al.* Super plastic bulk metallic glasses at room temperature. *Science* **315**, 1385–1388 (2007).
20. Lou, H. B. *et al.* 73 mm-diameter bulk metallic glass rod by copper mould casting. *Applied Physics Letters* **99**, 051910 (2011).
21. Xu, T., Pang, S., Li, H. & Zhang, T. Corrosion resistant Cr-based bulk metallic glasses with high strength and hardness. *Journal of Non-Crystalline Solids* **410**, 20–25 (2015).
22. Liu, Y. H. *et al.* Deformation behaviors and mechanism of Ni–Co–Nb–Ta bulk metallic glasses

with high strength and plasticity. *J. Mater. Res.* **22**, 869–875 (2007).
23. Meng, D. *et al.* Tantalum based bulk metallic glasses. *Journal of Non-Crystalline Solids* **357**, 1787–1790 (2011).
24. Bi, J. *et al.* OsCo-based high-temperature bulk metallic glasses with robust mechanical properties. *Scripta Materialia* **228**, 115336 (2023).
25. Zhou, Y. & Wang, T. High stability and high corrosion resistance of a class of Co–Cr–Mo–Nb–B high-entropy metallic glasses. *Journal of Materials Research and Technology* **30**, 256–266 (2024).